# Creating a Baseline for Robust Energy Savings Estimation in Households

Eftychios Protopapadakis, National Technical University of Athens, Greece, eftprot@mail.ntua.gr

Anastasia Garbi, European Dynamics, Belgium, anastasia.garbi@eurodyn.com

Anna Malamou, European Dynamics, Athens, anna.malamou@eurodyn.com

Maria Kaselimi, National Technical University of Athens, Greece, mkaselimi@mail.ntua.gr

Zisis Pontikas, European Dynamics, Athens, zisis.pontikas@eurodyn.com

Anastasios Doulamis, National Technical University of Athens, Greece, adoulam@cs.ntua.gr

Nikolaos Doulamis, National Technical University of Athens, Greece, ndoulam@cs.ntua.gr

Kostas Vasilakis, European Dynamics, Athens, kostas.vasilakis@eurodyn.com

Nassos Michas, European Dynamics, Athens, nassos.michas@eurodyn.com

Emmanouel Alexakis, National Technical University of Athens, Greece, alexmantud@gmail.com

**Abstract** In this paper we present a methodology appropriate for establishing a user-specific hourly-based benchmark period of energy consumption. Such values can be used as reference for explicitly calculating energy savings. The required equipment is limited to low cost sensors. The so-called baseline allows for an explicit comparison of the household consumption now and then, once appropriate adjustments are made to handle the different conditions (e.g. temperature variation). When involving user motivation strategies, hourly based energy savings can provide multiple advantages. However, there are two major factors affecting the baseline construction: corrupted or missing data values and unordinary patterns. In this paper we provide an analytical methodology to handle such scenarios and create year-long hourly-based consumption baseline.

**Keywords:** Energy savings, Pattern analysis, Consumption monitoring

## 1 Introduction

Increased awareness on environmental sustainability in individuals is not limited only at a global scale. There are multiple individuals, i.e. householders, interested in contributing to sustainable living, by reducing their energy consumption. Interaction design is an important factor, which can support this trend. The focus on more sustainable products can be divided into two broad categories: (i) sustainability through design – influencing everyday decisions by design of technology, (ii) sustainability in design – taking sustainability into account in the material design of products. Our focus is on the first category; we employ widely available low-cost sensors to generate numeric values that can be easily interpreted by the user. Generally, consumers recognize that new energy-efficient technologies for household devices are only part of the solution. Sustainable living also requires a change in behavior in the way household devices are used. Consumers first need to become aware of their energy consumption, to change their behavior.

Through the years, various players (e.g. companies, governments) made efforts at different levels (i.e. local, regional, national) to raise awareness about energy efficiency [1]. Nevertheless, end-consumers have limited knowledge about the existing policy measures and technologies available to them. Therefore, users fail to grasp the economic potential of energy savings, which consist a great motivation factor. Most of the household consumers are usually aware of general information related to their consumption, through monthly electricity bills. Nonetheless, the information about energy consumption is not translated into good practices and customized advices to save energy.

Up to this day, quantifying the results of energy efficiency is an extremely important task. A straightforward approach would be the direct comparison between a benchmark period, denoted as baseline, and the current consumption values. Typically, adjustments should be included. Quantification can be annually, monthly, daily, hourly, etc. However, denser the time interval higher the probability of missing a value, e.g. due to equipment failure). Establishing a baseline is not trivial. The most import problem is that users are not easily motivated in installing equipment in their houses, for a long time, without a direct beneficial impact to their routine. Applying non-intrusive load monitoring techniques [2], [3], [4], [5] for energy consumption monitoring and at the same time creating an accurate baseline model for energy savings estimation helps in the adoption of a personalized model for each user's energy consumption behavior.

In this paper we focus on establishing an accurate baseline by coupling consumption recordings, pattern analysis, data crawling from online resources and statistics from local authorities. The proposed approach can generate hourly-based year-long consumption patterns for any user, using two months or less of recorded values. Random consumption values can be generated following user specific distribution patterns and, consequently, be used for motivating towards a more energy-saving behavior. Described techniques were implemented during the BENEFFICE project [6].

## 2  Proposed methodology

The International Performance Measurement and Verification Protocol (IPMVP®) [7] defines standard terms and suggests practices for quantifying the results of energy efficiency investments. A straightforward approach to calculate energy savings is by comparison; check the difference between the baseline consumption and the current one. Therefore, creating an energy consumption baseline is a necessary, multistep, hierarchical process when we plan to calculate energy savings. Thus, a savings calculation process consists of three main phases: a) energy monitoring equipment installation/utilization, b) data handling and preprocessing, and c) calculating the savings. Steps a) and b) focus on the creation of the baseline. This work provides further details on the topic.

At first, we need to monitor/capture the energy consumption. However, consumption alone does not suffice; outside temperature is extremely important for reasons explained in section3. In this case we utilized a Raspberry Pi solution, using open source software, since it is highly parameterizable, replaceable and can be used with different power meters / sensors. Then, recorded data values are stored at a server, together with weather data. Stored values are subject to data handling and post-processing techniques, since corrupted or missing data is likely to exist. The last step involves the savings calculation.

### 2.1 Employed equipment

The sensing components consist of a) an energy consumption meter, placed on the central electrical board of the residence, and b) a temperature sensor to measure the indoor temperature of the residence. Collecting the readings of the smart meters/sensors and sending them to the BENEFFICE platform Backend, a device that will act as a gateway between the sensors and platform backend is needed. The use of the gateway device will ensure that the local readings of the meters and sensors are collected and forwarded safely and securely to the platform backend. Figure 1(a) demonstrates the employed data capturing concept.

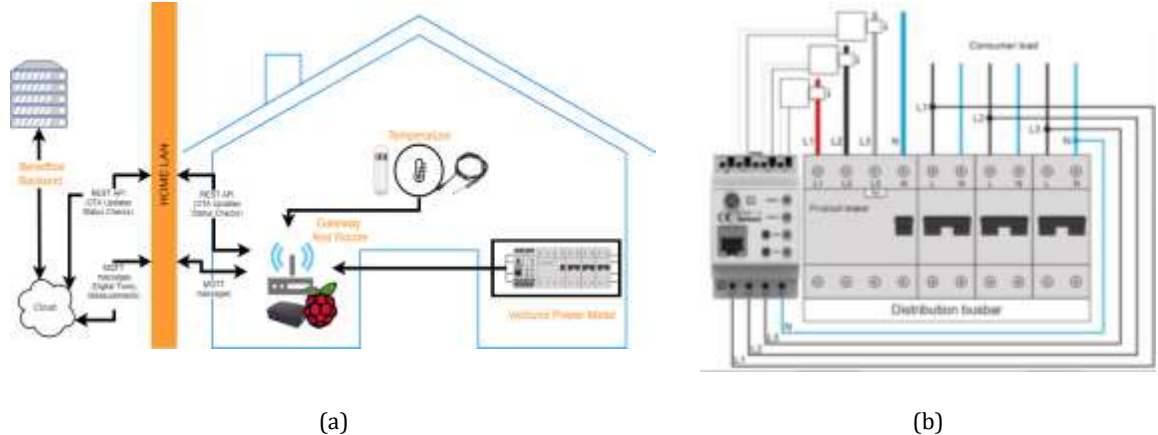

(a)  (b)

**Figure 1: Consumption data capturing concept. All the information from the sensors pass through house's local network to the gateway and then to the backend system. Image on the right demonstrates the power meter connectivity.**

The Energy consumption meter will be used to measure the total electricity consumption of the residence. Verbund power meter [8], a three-phase bi-directional power meter for electricity (see Figure 1(b)), was the device of choice. Communication with the gateway device can be established over the user's home network via powerline (PLC, additional adapter needed), WiFi or Ethernet cable. A Raspberry Pi [9] is used as a gateway device for the hardware part, while the openHAB platform [10] is used for the software part.

## 2.2 Data handling

Assume a sequence of hourly consumption values for a weekly period, $\boldsymbol{c}^{(w)}$, so that:

$$\boldsymbol{c}^{(w)} = \left[ \{c_i^{(M)}\}_{i=1}^{24}, \{c_i^{(Tu)}\}_{i=1}^{24}, \{c_i^{(W)}\}_{i=1}^{24}, \{c_i^{(Th)}\}_{i=1}^{24}, \{c_i^{(F)}\}_{i=1}^{24}, \{c_i^{(Sa)}\}_{i=1}^{24}, \{c_i^{(Su)}\}_{i=1}^{24} \right] \quad (1)$$

where, $\{c_i^{(k)}\}_{i=0}^{24} = [c_1^{(k)}, \dots, c_{24}^{(k)}]$, $k = \{M, Tu, W, Th, F, Sa, Su\}$ denotes the hourly consumption values at hour $i$ the day $k$. The energy used, by the household for a year, $E_b^{(y)}$, is then provided by the following equation:

$$E_b^{(y)} = \sum_{w=1}^{52} \boldsymbol{c}^{(w)} \quad (2)$$

At this point, we can understand that obtaining hourly consumption values for a year-long period is a difficult task. On the one hand, missing values due to equipment or software failure and human errors is a common case. The main problems observed under this category were: a) VPM and gateway connectivity failure (due to low Wi-Fi signal strength close to the power board), b) accidental unplugging of the gateway from the power supply, c) hardware failure (SD card) on the gateway. On the other hand, the duration of the period is a significant constraint. Typically, a user will not accept any installations within the house for such a long period, without any practical feedback or short-term benefits. Considering the above, a significant amount of information will not be available to create a baseline period.

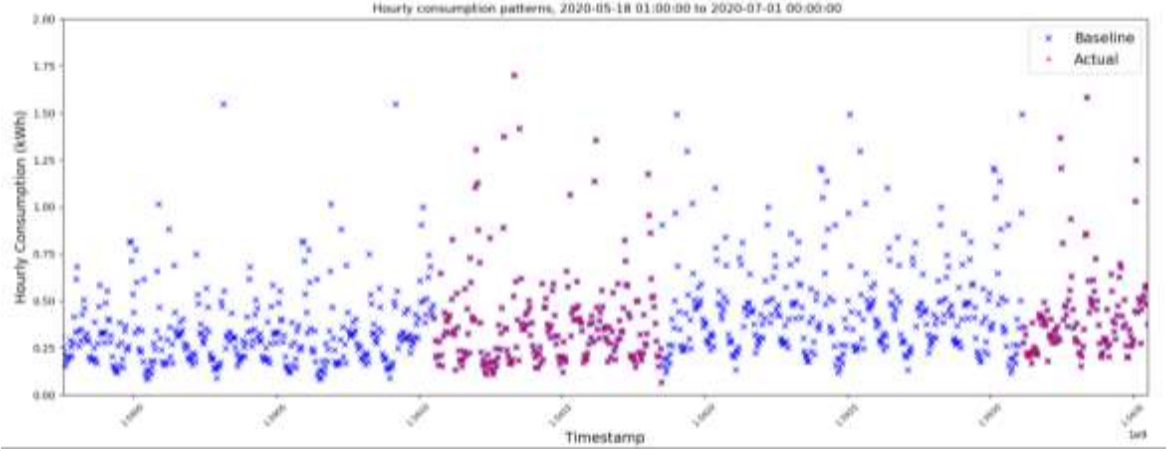

**Figure 2:** Illustrating the generated missing consumption values, given a set of observations, for a specific user.

Let as assume a set of monthly consumption values, $\hat{c}^{(m)} = \{\hat{c}_i^{(m)}\}_{i=1}^{12}$, provided by Eurostat [11], the statistical office of the European Union. These values need to be adjusted and distributed over an hourly base. At first, we will proceed with the adjustment factor, $n_f$, which is defined as:

$$n_f = \frac{1}{k}\sum_{i=1}^{k}\frac{\hat{c}_i^{(m)} - c_i^{(m)}}{\hat{c}_i^{(m)}} \quad (3)$$

where $\hat{c}_i^{(m)}$ and $c_i^{(m)}$ are the aggregated monthly consumption values, $i = 1, \ldots, k, k \leq 12$ for all the months that we have hourly consumption recordings. Practically, $k \in \{1,2,3\}$, since most of the users had two months of recorded observations, obtained using the process described in section 2.2. Once the adjustment factor $n_f$ is calculated the monthly energy consumption per user can be expressed as: $c^{(m,u)} = \{\tilde{c}_i^{(m,u)}\}_{i=1}^{12}$, where $\tilde{c}^{(m,u)}$ is defined as:

$$\tilde{c}_i^{(m,u)} = \begin{cases} c_i^{(m)}, & iff\ c_i^{(m)}\ was\ recorded \\ n_f \cdot \hat{c}_i^{(m)}, & else \end{cases} \quad (4)$$

The last step involves a mapping process $f(\cdot): \hat{c}_i^{(m)} \rightarrow c^{(w)}$ so as to create the hourly consumption values. Using the available (observed) hourly consumption values we can create a weekly percentage distribution:

$$\boldsymbol{p}^{(w)} = \left[\{p_i^{(M)}\}_{i=1}^{24}, \{p_i^{(Tu)}\}_{i=1}^{24}, \{p_i^{(W)}\}_{i=1}^{24}, \{p_i^{(Th)}\}_{i=1}^{24}, \{p_i^{(F)}\}_{i=1}^{24}, \{p_i^{(Sa)}\}_{i=1}^{24}, \{p_i^{(Su)}\}_{i=1}^{24}\right] \quad (5)$$

where $p_i^{(d)} = c_i^{(d)}/\sum c^{(w)}, i = 1, \ldots, 24$ and $d = \{M, Tu, W, Th, F, Sa, Su\}$. Then, the $i$-th hourly consumption value, for day $d$, on month $m$, for a user $u$, denoted as $c_i^{(d,m,u)}$, can be calculated as:

$$c_i^{(d,m,u)} = p_i^{(d)} \cdot \tilde{c}_k^{(m,u)}/4 \quad (6)$$

where $i = 1, \ldots, 24$ and $k = 1, \ldots, 12$. Recall that there are 4 weeks per month. In this approach, $p_i^{(d)}$ vales are constant for every week and every month. Yet, $\tilde{c}_k^{(m,u)}$ varies allowing for better consumption estimations. Figure 2 demonstrates the baseline creation, i.e. establishing a user specific benchmark period. Points denoted with red crosses (+) correspond to measurements obtained using the proposed equipment. According to these points, we were able to formulate the expected hourly consumption distribution percentages. Points denoted as

(x) are randomly generated consumption patterns, which follow the same distribution of the observed ones, adjusted accordingly to the month we focus.

## 2.3 IPMVP protocol

Energy savings are determined by comparing measured energy use before and after implementation of an energy savings program or related actions. The equation can be stated as:

$$S = E_b - E_{pr} \pm A \qquad (7)$$

where, $S$ denotes the energy savings (kWh), $E_b$ and $E_{pr}$ correspond to baseline and post-retrofit energy use, and $A$ indicates an adjustment term, which brings energy use in the two time periods to the same set of conditions. Conditions commonly affecting energy use are weather, occupancy, plant throughput, and equipment operations required by these conditions. Adjustments may be positive or negative.

## 3 Calculating the energy savings

At this point we should note that seasonal trends (i.e. temperature fluctuations) affect the trend calculation. There are multiple techniques suitable for such cases, e.g. Degree Days Temperature methodology (DDT) [12]. DDT method can track energy usage, as it allows the comparison of energy consumption data between different periods of time, limiting the temperature bias. Yet, simplified concepts can be used.

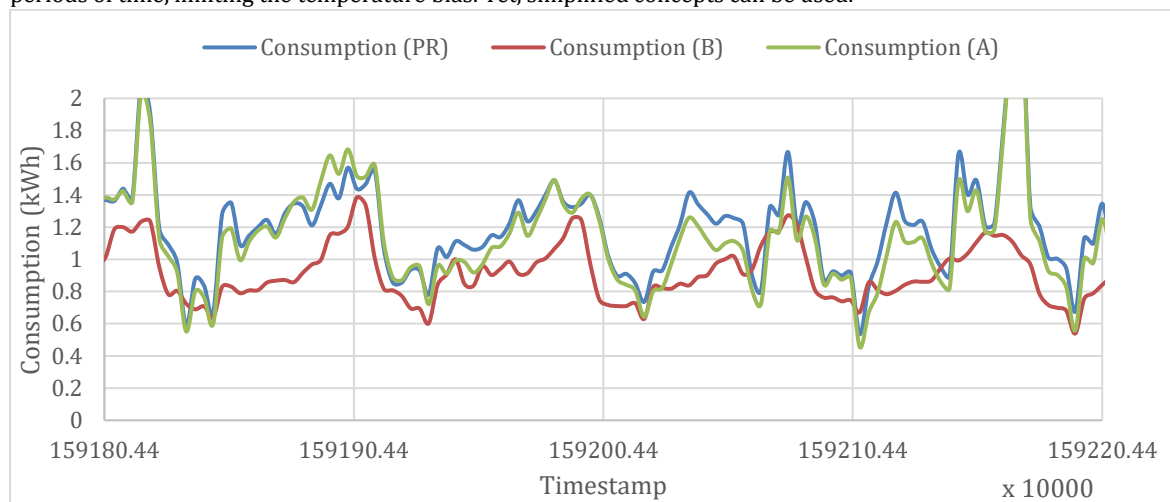

**Figure 3:** Demonstrating the Consumption values (baseline, observed, and adjusted)

Figure 3 illustrates the proposed approach for a period of 5 days in June, starting at 2020,06,10,19:00 (or 1591804800 as a timestamp), ending at 2020,06,15,10:00 (or 1592204400 as a timestamp). The user location was Athens, Greece, with an average temperature of 24.1 °C, for the observed period. The baseline period had an average temperature of 25.79 °C. We start by noting the current consumption values, prior to any adjustments (PR). Then, the observed hourly values, denoted with (A), are adjusted by a factor equal to the ratio of current temperature to baseline equivalent, creating a new line.

## 4 Conclusions

A detailed approach suitable for generating hourly-based, year-long benchmark period of energy consumption values, capable to support detailed energy savings calculation is proposed. The adopted methodology is user specific, has no significant installation/ implementation costs and can work with approximate two month of actual observations. The missing values can be handled appropriate, using country specific statistics and pattern analysis.

**ACKNOWLEDGMENTS**

The EU H2020 BENEFFICE project "Energy Behaviour Change driven by plug-and play-and-forget ICT and Business Models focusing on complementary currency for Energy Efficiency for the Wider Population" has received funding from the European Union's Horizon 2020 research and innovation programme under grant agreement No 768774. A researcher was also funded by General Secretaruy of Research and Development (Greek GSRT), project ID 671168.

7